\documentclass[a4paper,fleqn,usenatbib]{mnras}

\usepackage[T1]{fontenc}
\usepackage{ae,aecompl}

\usepackage{graphicx}	
\usepackage{amsmath}	
\usepackage{amssymb}	
\usepackage{multicol}        
\usepackage{bm}		
\usepackage{color}
\usepackage{xcolor}

\title[Pulsation among {\it TESS} A and B stars]
{Pulsation among {\it TESS} A and B stars and the Maia variables}
\author[L. A. Balona and D. Ozuyar]{\
L. A. Balona$^{1}$\thanks{E-mail: lab@saao.ac.za} and
D. Ozuyar$^{2}$
\\
$^1$ South African Astronomical Observatory, 
P.O. Box 9, Observatory 7935, South Africa\\
$^2$ Ankara University, Faculty of Science, 
Dept. of Astronomy and Space Sciences, 06100, Tandogan, Ankara, Turkey}

\begin{document}

\date{Accepted .... Received ...}

\pagerange{\pageref{firstpage}--\pageref{lastpage}} \pubyear{2011}

\maketitle

\label{firstpage}

\begin{abstract}
Classification of over 50000 {\it TESS} stars in sectors 1--18 has resulted
in the detection of 766 pulsating main sequence B stars as well as over
5000 $\delta$~Scuti, 2300 $\gamma$~Doradus and 114 roAp candidates.  Whereas
it has been assumed that high frequency pulsation among B-type main sequence 
stars are confined to the early B-type $\beta$~Cephei stars, the
observations indicate that high frequencies are to be found over the whole 
B-star range, eventually merging with $\delta$~Scuti stars.  The cool B stars 
pulsating in high frequencies are called  Maia variables.  It is shown that 
Maia variables are not rapidly-rotating and thus cannot be $\beta$~Cephei
pulsators which appear to have lower temperatures due to gravity darkening.
In the region where $\beta$~Cephei variables are found, the proportion of 
pulsating stars is larger and amplitudes are higher and a considerable 
fraction pulsate in a single mode and low rotation rate. There is no distinct 
region of slowly-pulsating B stars (SPB stars).  Stars pulsating solely in low 
frequencies are found among all B stars.  At most, only one-third of B stars 
appear to pulsate.  These results, as well as the fact that a large fraction of
A and B stars show rotational modulation, indicate a need for a revision of 
current ideas regarding stars with radiative envelopes.
\end{abstract}


\section{Introduction}

The advent of space photometry, particularly from the {\it Kepler} and {\it
TESS} missions, has radically changed perceptions on stellar pulsation in
the upper main sequence.  For example, from {\it CoRoT} data, 
\citet{Degroote2009b} find evidence for a new class of low-amplitude B-type 
pulsators between the SPB and $\delta$~Scuti instability strips.
Prior to {\it Kepler}, it was believed that the opacity $\kappa$ mechanism 
offered a complete and satisfactory explanation for the $\delta$~Scuti, SPB 
and $\beta$~Cephei pulsating variables. At the cool end of the $\delta$~Sct 
instability strip, pulsations in the $\gamma$~Doradus variables were 
attributed to the convective blocking mechanism \citep{Guzik2000}.  Photometric time 
series observations from space were expected to confirm model predictions and 
perhaps resolve a few minor problems.

The first surprise was the discovery that low-frequency $\gamma$~Dor pulsations
are visible throughout the $\delta$~Sct instability strip \citep{Grigahcene2010}.  
We know now that the distinction between the two types of variable is merely
one of mode selection and not of pulsation mechanism since they share the
same instability region \citep{Balona2018c}.  This suggests a problem with
assumptions regarding convection on the upper main sequence. \citet{Bowman2018}
and \citet{Balona2018c} find that a non-negligible fraction of main-sequence 
$\delta$~Sct stars exist outside theoretical predictions of the classical 
instability boundaries.  Recent calculations using time-dependent perturbation 
theory \citep{Antoci2014b,Xiong2016} including turbulent convection do not 
resolve the issue.

Among the main-sequence B stars, the $\kappa$~mechanism operating in the
ionization zone of iron-group elements appears to be responsible for the
high-frequency pulsations of the $\beta$~Cephei stars as well as the
low-frequency pulsations among the cooler SPB variables.  These were the
only two recognized classes of pulsating variable in main-sequence stars 
hotter than the blue edge of the $\delta$~Sct instability strip.  
\citet{Stankov2005} provide a list of 93 confirmed $\beta$~Cep variables
with an additional 14 stars discovered by \citet{Pigulski2005}.  Most recently,
\citet{LabadieBartz2019} presented results of a search for $\beta$~Cep stars 
from the {\it KELT} exoplanet survey.  They identify 113 $\beta$~Cep stars, of 
which 86 are new discoveries.  \citet{Burssens2019} found 3 new $\beta$~Cep 
stars observed by the {\it K2} mission.  Ground-based surveys for SPB stars 
have been made by several groups \citep{Aerts1999, Mathias2001, deCat2002, 
deCat2007}.  

The {\it CoRoT} space mission \citep{Fridlund2006} contributed considerably to 
our knowledge of pulsation among the B stars.  For example,  observations of 
the O9V star HD\,46202 show $\beta$~Cep-like pulsations, but none of the 
observed frequencies are excited in the models \citep{Briquet2011}.  A global 
magnetic field was found in the  hybrid B-type pulsator HD\,43317 
\citep{Briquet2013}.  Unexpected modes with short lifetimes in HD\,180642 were 
initially interpreted as stochastic modes excited by turbulent convection 
\citep{Belkacem2009}, but this conclusion was subsequently disputed 
\citep{Aerts2011, Degroote2013}.  At least 15 new SPB candidates were detected 
by {\it CoRoT} \citep{Degroote2009b}.

Due to the high galactic latitude of the field observed by the {\it Kepler} 
space mission \citep{Borucki2010}, relatively few B stars were observed. 
\citet{Balona2011b} found 15 pulsating stars, all of which show low 
frequencies characteristic of SPB stars. Seven of these stars also show a few 
weak, isolated high frequencies.

From time to time, ground-based observations reported the possible presence of 
high frequencies in a few stars too hot to be $\delta$~Scuti, but too cool
to be $\beta$~Cephei \citep{McNamara1985,Lehmann1995,Percy2000, Kallinger2004}.
These  were called Maia variables following a report by \citet{Struve1955} of 
short-period variations in the star Maia, a member of the Pleiades cluster. 
\citet{Struve1955} later disclaimed the variability. It is now known from
the {\it K2} space mission that Maia itself is a rotational variable with a 
10-d period \citep{White2017} and no sign of high frequencies.  

One of the most interesting results from the {\it CoRoT} mission is evidence 
for a new class of low-amplitude B-type pulsator between the SPB and 
$\delta$~Sct instability  strips, with a very broad range of frequencies
extending well into the $\beta$~Cep range \citep{Degroote2009b}.  These are 
probably the same as the Maia variables described above.  B stars with high
frequencies too cool to be $\beta$~Cep variables have also been detected in
the {\it Kepler} field \citep{Balona2015c, Balona2016c}.

From ground-based photometry, \citet{Mowlavi2013, Lata2014, Mowlavi2016}
discovered anomalous high frequencies in rapidly-rotating late to
mid-B stars.  At this time, it is not clear if these stars can
be considered as Maia variables. From pulsating models of rotating B stars,
\citet{Salmon2014} found that frequencies as high as 10\,d$^{-1}$ may be
visible in mid- to late B stars.  They therefore suggest that these 
could be fast-rotating SPB stars in which the apparent effective temperature
is lowered by gravity darkening at the equator.  More recently, 
\citet{Szewczuk2017} have computed the instability domains for B stars 
including the effects of slow-to-moderate rotation.   They find that unstable 
prograde high radial order g modes may have quite high frequencies which could
account for these anomalous variables and the Maia stars.
 
The advent of {\it TESS} \citep{Ricker2015} has opened a new opportunity to 
study pulsation among the B stars.  A sample of over 50000 stars hotter than 
6000\,K from sectors 1--18 were examined and classified according to variability 
type.  The classification is based on a visual inspection of the periodogram 
and light curve for each star.  According to current knowledge, apart from the
chemically peculiar roAp stars,  the only A and B main sequence stars with 
frequencies higher than about 5\,d$^{-1}$ are the $\delta$~Sct (DSCT) and 
$\beta$~Cep (BCEP) variables, the distinction being made according to the  
effective temperature.  It is, however, necessary to introduce the Maia class 
to account for the anomalous high-frequency B-type variables just described.  
This does not imply that the Maia variables are necessarily a separate group of
pulsating stars.  If Maia variables are simply rapidly-rotating $\beta$~Cep or 
SPB stars, then they must have higher than normal projected rotational 
velocities, allowing a test of this idea to be made.  

The effective temperatures, $T_{\rm eff}$, in the {\it TESS Input Catalogue} 
(TIC) \citep{Stassun2018} are unreliable for B stars because most are derived 
from multicolour photometry lacking the U band.  Without photometric 
measurements in the U band, the Balmer jump cannot be measured and it is not 
possible to distinguish stars with $T_{\rm eff} > 10000$\,K from the A stars.  
Spectral classification shows that there are nearly 2900 B stars in the sample 
of about 50000 {\it TESS} stars.  Many stars that we originally classified as 
DSCT or $\gamma$~Doradus (GDOR) on the basis of temperatures listed in the TIC, 
were re-classified as BCEP, MAIA or SPB variables because they have B-type 
spectra.

In this paper, we use the best available estimates of $T_{\rm eff}$ 
and luminosities using {\it Gaia DR2} parallaxes to locate the stars in the 
H--R diagram and the pulsation period vs $T_{\rm eff}$ (P--T) diagram.  The 
locations of stars classified as SPB or BCEP in the P--T diagram  are compared 
to the predicted locations derived from pulsation models. In this way, current 
ideas regarding pulsation instability among the B stars can be tested.  In 
particular, we investigate the status of the Maia variables and the connection 
between pulsation in A and B stars.

\section{Data and classification}

{\it TESS} light curves for thousands of stars with two-minute cadence are
available according to sector number.  There are 26 partially overlapping 
sectors covering the whole sky and each sector is observed for approximately 
one month.  The wide-band photometry has been corrected for long-term drifts
using pre-search data conditioning  (PDC, \citealt{Jenkins2010b}).  Each 
{\it TESS} pixel is 21\,arcsec in size which is similar to the typical aperture
size used in ground-based photoelectric photometry.  Working groups 4 and 5 of 
the {\it TESS} asteroseismic consortium were involved in target selection
\citep{Pedersen2019, Handler2019}.

Effective temperatures are listed for most {\it TESS} stars in the TIC, but
for reasons already mentioned, they cannot be used for B stars.  To resolve
this problem,  a catalog of over 600000 stars brighter than 12-th magnitude
with known spectroscopic classifications was created and matched with the
TIC.  This allows the proper assignment to be made regarding the type of
variability.  

As far as possible, it is necessary to assign the variability 
class in accordance with the well-established types used in the 
{\it General Catalogue of Variable Stars} (GCVS, \citealt{Samus2017}).
One major consideration is that variation due to rotational modulation seems
to be present in all types of star, including the B stars \citep{Balona2019c}.
This may be due to chemical peculiarities (the SX~Ari class), but most often 
there is no indication of spectral peculiarity, in which case the new ROT class
is assigned.  Rotation always needs to be considered when examining the 
periodograms at low frequencies.  

Variability classification for stars with $T_{\rm eff} > 6000$\,K is made
as each sector becomes available.  Aided by suitable software, classification 
of over 100 stars an hour is possible, so that the full sector, which normally 
contains 1000--2000 previously unclassified stars, can be classified in a few 
days.  In this way, variability types of over 50000 stars in sectors 1--18 were 
obtained.  Of these, 2868 are O and B stars.  An important advantage of visual 
classification is that after a while it is easy to recognize early-type 
supergiants, Be stars, and many other classes purely from the light curve and 
periodogram.  Of course, many interesting stars can be noted for further study.

As already mentioned, pulsation models of non-rotating $\beta$~Cep and SPB 
stars show a clear separation between the two kinds of variable in a plot of
pulsation period as a function of $T_{\rm eff}$ (see Fig.\,4 of
\citealt{Miglio2007b}).  SPB stars mostly have frequencies less than 
2.5\,d$^{-1}$, whereas $\beta$~Cep stars all have higher frequencies.  Most 
$\beta$~Cep stars are confined to $T_{\rm eff} > 18000$\,K. We therefore 
classified a star as BCEP only if it is earlier than B5 or B6 and if most of 
the observed frequencies of highest amplitude exceed 2.5\,d$^{-1}$.  For SPB 
stars it was decided that for $T_{\rm eff} > 18000$\,K a boundary of 
2.5\,d$^{-1}$ separates BCEP from SPB, but for cooler stars the boundary 
between low and high frequencies was moved to 5\,d$^{-1}$ to allow more 
flexibility for rotational effects.  A star is classified as MAIA if 
$10000 < T_{\rm eff} < 18000$\,K and most peaks of high amplitude exceed 
5\,d$^{-1}$.

In many cases the amplitudes in the low-frequency range are comparable to those
in the high frequencies in which case we use BCEP+SPB or SPB+BCEP, depending on
which range seems to dominate.  These would be the $\beta$~Cep/SPB hybrids.
The hybrid classes MAIA+SPB and SPB+MAIA are also assigned.

\begin{figure}
\begin{center}
\includegraphics{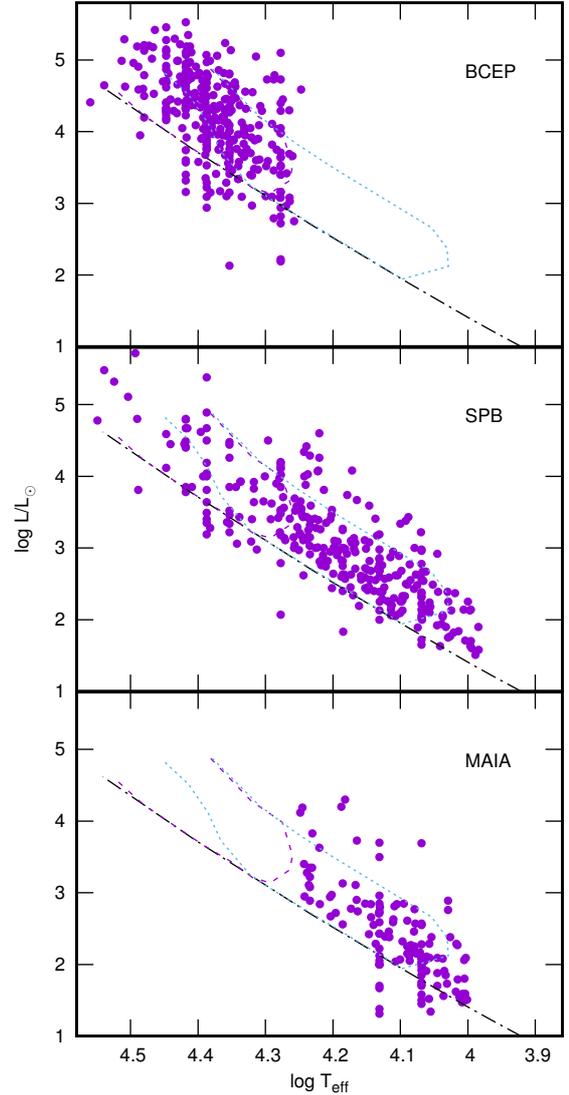}
\caption{The H--R diagram showing the $\beta$~Cep, SPB and Maia stars observed 
by {\it TESS}. Also shown is the theoretical zero-age main sequence (solid 
line) and the instability regions of the $\beta$~Cep and SPB pulsating stars
for $Z = 0.02$ and spherical harmonic degree $l \le 3$ from
\citet{Miglio2007}.}
\label{hrdiag}
\end{center}
\end{figure}

In Table\,\ref{bcep}, 327 $\beta$~Cep variables (the majority of which
are previously unknown), 308 pure SPB stars (i.e. non-hybrids) and 131 Maia 
stars are listed.  In Fig.\,\ref{hrdiag}, the $\beta$~Cep, SPB and Maia stars 
are shown in the theoretical H-R diagram.  Also shown are the theoretical 
instability strips for metal abundance $Z = 0.02$, using OP opacities and the 
AGS05 mixture as calculated by \citet{Miglio2007b}.  A more recent calculation 
of the instability region by \citet{Walczak2015} using updated opacities gives 
much the same result.  A few stars lie below the zero-age main sequence and may
be previously unrecognized pulsating subdwarfs.  

It should be noted that the calculated instability strips shrink quite rapidly 
with decreasing metallicity.  The figure shows the instability strips for solar 
abundance and OP opacities from \citet{Miglio2007b}.  These define approximately 
the maximum extent of instability in non-rotating models.  Rotation will
tend to reduce the apparent effective temperature of a star if it is rapidly
rotating and with high inclination due to gravity darkening at the equator
\citep{VonZeipel1924}.  This means that some stars may appear outside the
cool edge of the instability strip.  \citet{Szewczuk2017} computed the 
instability domains for gravity and mixed gravity-Rossby modes, including the 
effects of slow-to-moderate rotation.  The main result is that g-mode 
instability domains are much more extended towards higher masses and higher 
effective temperatures, mainly as a result of using OPAL rather than OP 
opacities.

\begin{table}
\caption{List of $\beta$~Cep, SPB and Maia stars in {\it TESS} sectors 1--18.  
The priority code of effective temperature, $T_{\rm eff}$, is given in column
4 as follows: 1 - spectroscopy; 2 - narrow-band photometry; 3 - UBV photometry;
4 - BV photometry; 5 - Spectral type or other.  The complete table is available
in electronic form.}
\label{bcep}
\begin{center}
\resizebox{8cm}{!}{
\begin{tabular}{rlrrrrl}
\hline
\multicolumn{1}{c}{TIC} & 
\multicolumn{1}{c}{Var Type} & 
\multicolumn{1}{c}{$T_{\rm eff}$} & 
\multicolumn{1}{c}{Pr} & 
\multicolumn{1}{c}{$\log\tfrac{L}{L_\odot}$} & 
\multicolumn{1}{c}{$v\sin i$} & 
\multicolumn{1}{l}{Sp. Type}  \\
\hline
    99010 & SPB      &  9931 & 1 & 2.13 & 164 & B9.5III      \\
  3300381 & BCEP+ROT & 20400 & 3 & 4.09 &     & B2IIIn       \\
  4207261 & SPB+BCEP & 24012 & 1 & 4.03 &  10 & B1.5V        \\
  6110321 & MAIA+SPB & 14200 & 4 & 2.85 &     & B8e:         \\
  7429754 & SPB      & 15546 & 2 & 2.96 &     & B6V          \\
  9887122 & SPB+MAIA & 11324 & 1 & 2.38 & 150 & B6(V)        \\
 10510382 & SPB      & 22570 & 5 & 3.85 &  37 & B3Vp shell   \\
 10891640 & BCEP+EB  & 28000 & 5 & 4.29 &     & B0.5III:     \\
 11400562 & SPB+BCEP & 23940 & 1 & 4.13 &  75 & B2IV-V       \\
 11411724 & BCEP+SPB & 24068 & 2 & 3.15 & 200 & B1.5V        \\
 11696250 & BCEP     & 26169 & 1 & 5.07 & 116 & B0.5III      \\
 11698190 & BCEP     & 20200 & 3 & 4.29 &     & B0.5V        \\
\hline                        
\end{tabular}
}
\end{center}
\end{table}

\section{Effective temperatures and luminosities}

The effective temperature is a crucial component in establishing
the type of variability.  The most reliable estimates of $T_{\rm eff}$ are
those which use spectroscopic observations combined with model atmospheres.
The {\it PASTEL} compilation of spectroscopic parameters \citep{Soubiran2016} 
are particularly useful in this regard.  The literature was searched for
more recent measurements, and a catalogue of over 101500 stars comprising
nearly 170000 individual $T_{\rm eff}$ measurements of various kinds was
compiled.  

Each method of deriving $T_{\rm eff}$ was assigned a priority
class.  First priority is given to spectroscopic modelling.  Values of
$T_{\rm eff}$ from Str\"{o}mgren or Geneva photometry were assigned second
priority.  Values using Johnson UBV photometry and the Q method of
de-reddening were applied to many stars, but assigned third priority.
Fourth priority was given when only BV photometry is available.  In this
case the reddening can be estimated using a 3D reddening map by 
\citet{Gontcharov2016}.  The \citet{Torres2010b} calibration giving 
$T_{\rm eff}$ as a function of $(B-V)_0$ was used for the latter two methods.  
Finally, when no other way of obtaining $T_{\rm eff}$ was possible, the
spectral type and luminosity class and the \citet{Pecaut2013} calibration
were used and assigned fifth priority.   The adopted value of $T_{\rm eff}$
is the average of measurements of highest priority only, even if many more
measurements of lower priority are available.  For most stars, only one
measurement method (i.e. one priority) is available.

Stellar luminosities were derived from  {\it Gaia DR2} parallaxes 
\citep{Gaia2016,Gaia2018}.  The bolometric correction was obtained from 
$T_{\rm eff}$ using the \citet{Pecaut2013} calibration.  The reddening 
correction was derived from a three-dimensional reddening map  by 
\citet{Gontcharov2017}.   From the error in the {\it Gaia} DR2 parallax, the 
typical standard deviation in $\log(L/L_\odot)$ is estimated to be about 
0.05\,dex, allowing for standard deviations of 0.01\,mag in the apparent 
magnitude, 0.10\,mag in visual extinction and 0.02\,mag in the bolometric 
correction in addition to the parallax error.

A catalogue of projected rotational velocities, $v\sin i$ consisting of
over 58000 individual measurements of 35200 stars was compiled.  The bulk 
of these measurements are from \citet{Glebocki2005b}.  The catalogue was
brought up to date by a literature search.

\begin{figure}
\begin{center}
\includegraphics{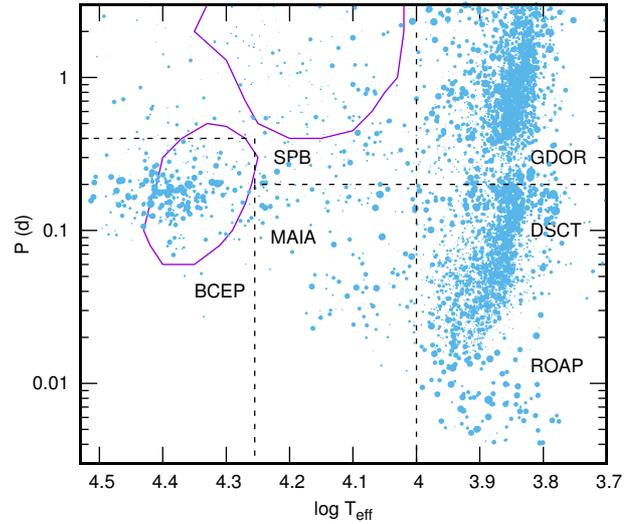}
\caption{Location of {\it TESS} pulsating stars stars in the period/effective 
temperature diagram.  For each star, the frequency of maximum amplitude is
plotted with size proportional to logarithm of the amplitude.  The two oval 
regions show the location of unstable modes of low degree calculated by 
\citet{Miglio2007}.  The regions demarcated by the dashed lines are the adopted
locations of the $\beta$~Cep, SPB, Maia, $\delta$~Sct and $\gamma$~Dor
stars.}  
\label{miglio}
\end{center}
\end{figure}

\section{The Period-Temperature diagram}

\citet{Miglio2007} studied the effect of different opacity tables and
metallicities on the pulsational stability of non-rotating B stars.  A useful 
visual representation of these results is a plot of the periods of unstable 
modes as a function of $T_{\rm eff}$ (the P-T diagram).  This leads to two 
non-overlapping regions, as shown in Fig.\,\ref{miglio},  which define the 
instability regions of $\beta$~Cep and SPB stars in the models.  One may
expect a larger spread of pulsation periods and a displacement in the observed 
$T_{\rm eff}$ to lower values in the more rapidly-rotating stars.
 
Also shown in Fig.\,\ref{miglio} are the pulsating stars observed by 
{\it TESS} where the symbol size is related to the maximum amplitude. 
To obtain the period in the P-T diagram, the peak of highest amplitude with 
frequency $\nu > 2.5$\,d$^{-1}$ was used for a star classified as BCEP.  For 
BCEP+SPB or SPB+BCEP, two periods are extracted, one above 2.5\,d$^{-1}$ and 
the other below this frequency.  For SPB, only one period with 
$\nu < 2.5$\,d$^{-1}$ is extracted.  For stars with $T_{\rm eff} < 18000$\,K, 
the cool edge of the $\beta$~Cep instability strip, one period with 
$\nu < 5$\,d$^{-1}$ was  extracted.  For SPB+MAIA or MAIA+SPB, two periods are 
extracted and for MAIA stars, the peak of highest amplitude with 
$\nu > 5$\,d$^{-1}$ is used.  For comparison, Fig.\,\ref{miglio} also shows
the DSCT, GDOR and ROAP stars.  The dashed lines in the figure represent the
regions for the different variability classes as defined above.

In Fig.\,\ref{miglio} there is a concentration of stars with relatively high 
amplitudes in the predicted $\beta$~Cep region (though somewhat displaced to 
higher $T_{\rm eff}$), mostly with frequencies in the range  5--10\,d$^{-1}$.  
There is no obvious concentration of SPB stars,  but there are quite a number 
of stars cooler than the $\beta$~Cep stars with high frequencies.  These 
are the Maia variables.

\begin{figure}
\begin{center}
\includegraphics{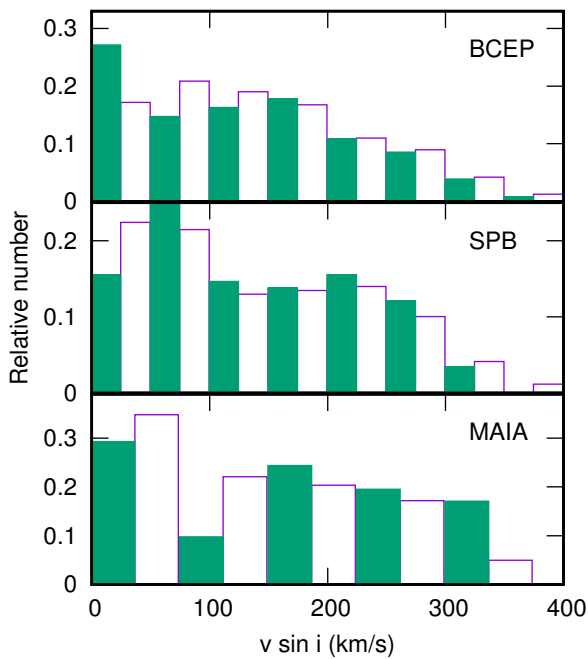}
\caption{The distribution of projected rotational velocities, $v\sin i$, for
main sequence stars (open boxes) and for $\beta$~Cep, SPB and Maia stars 
(filled boxes).}
\label{vdist}
\end{center}
\end{figure}

\section{$\beta$~Cep stars}

The frequency distribution of $\beta$~Cep stars, as revealed by the periodograms, is 
much sparser than that of $\delta$~Sct stars.  Very few $\beta$~Cep stars have 
more than a dozen significant peaks and about 20-30\,percent pulsate in just one 
dominant frequency.  Very often harmonics are present in these single-mode
stars.  The median amplitude of the maximum-amplitude peak for all
$\beta$~Cep stars is 3.9\,ppt, while for the $\beta$~Cep stars with one 
dominant mode it is 13.3\,ppt.  The average projected rotational velocity for 
all $\beta$~Cep stars is $\langle v\sin i\rangle = 131 \pm 8$\,km\,s$^{-1}$ 
(129 stars).  For stars with a single pulsation mode $\langle v\sin i\rangle = 
70\pm 15$\,km\,s$^{-1}$ (23 stars).  Thus single-mode $\beta$~Cep stars have 
significantly higher amplitudes and significantly lower rotation rates than 
other $\beta$~Cep stars.  The implication of this result is not clear at 
present.

For all B stars in the same temperature range $\langle v\sin i\rangle = 142 
\pm 2$\,km\,s$^{-1}$ (2730 stars).  The rotational frequency distribution for 
$\beta$~Cep stars is therefore the same as for all main sequence stars in
the same temperature range, as shown in Fig.\,\ref{vdist}.

It is reasonable to presume that the dominant high-amplitude mode could 
have the same spherical harmonic degree, $l$, in these stars.
The dimensionless frequency, $\sigma = \omega\sqrt{R^3/GM}$ where $\omega$
is the angular pulsation frequency, $R$ the stellar radius, $M$ the
stellar mass and $G$ the gravitational constant, is a useful indicator of
the pulsation mode.  Given $T_{\rm eff}$ and the stellar luminosity, $M$ can
be estimated from interpolation of evolutionary tracks which enables
$\sigma$ to be found. From 71 $\beta$~Cep stars having a single dominant 
frequency, it is found that $\langle\sigma\rangle = 5.4 \pm 0.3$.  The main 
source of error is the effective temperature.  An uncertainty of 1000\,K in 
$T_{\rm eff}$ leads to an uncertainty of 0.7 in $\sigma$.  Comparison with 
theoretical models calculated using the code by \citet{Dziembowski1977a} 
suggests that a radial mode $l = 0, n = 3$ or $l = 2, n = 3$, where $n$ 
is the radial order, as the most probable identification.

\begin{figure}
\begin{center}
\includegraphics{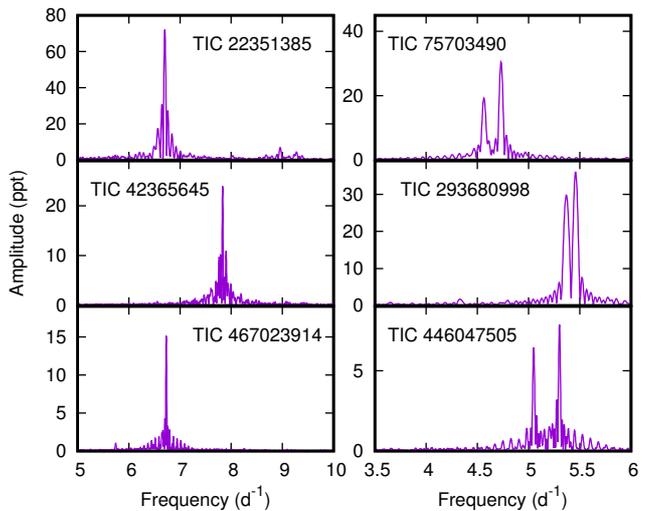}
\caption{Examples of $\beta$~Cep stars with just a single periodogram peak
(left hand panels) and with a single double peak (right hand panels).}
\label{bcepex}
\end{center}
\end{figure}

About 8 percent of $\beta$~Cep stars have a curious feature.  This is the presence of 
just two peaks of similar amplitude, perhaps a result of rotational splitting
or tidal interaction.  These stars are quite noticeable in visual inspection
of the periodograms and deserve careful study.  Examples are shown in 
Fig.\,\ref{bcepex}.  About half the $\beta$~Cep stars contain low frequencies 
typical of SPB stars, i.e. they are $\beta$~Cep/SPB hybrids.

\begin{figure*}
\begin{center}
\includegraphics{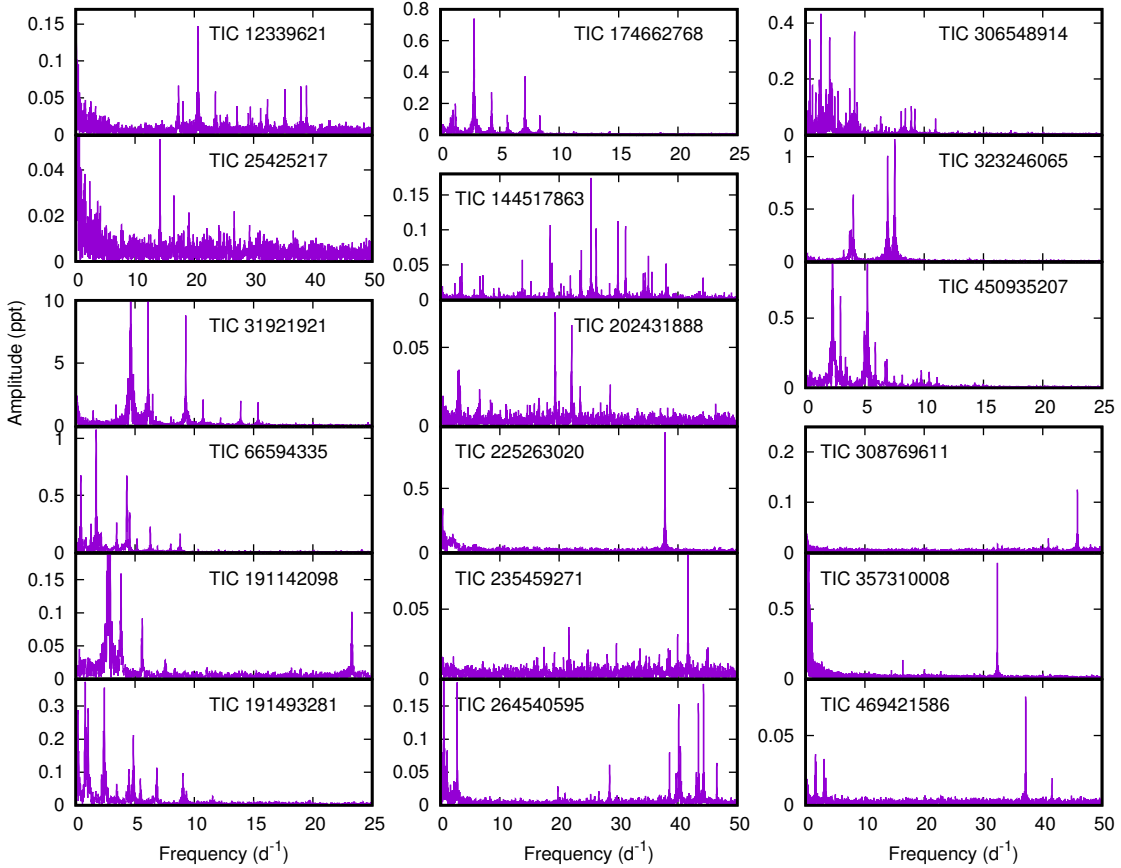}
\caption{Periodograms of some Maia stars in Table\,\ref{maiarot}.}
\label{pgmaia}
\end{center}
\end{figure*}

\section{SPB stars}

As can be seen in Fig.\,\ref{hrdiag}, stars with only low frequencies 
(i.e. the SPB stars with peaks below 2.5\,d$^{-1}$ or 5\,d$^{-1}$) are 
prevalent right across the B star region, including a substantial number within
the $\beta$~Cep instability  region.  Although hybrid stars are expected in 
this region, models do not predict stars where only low frequencies are 
unstable.  There are 90 pure SPB stars (i.e. no frequencies exceeding 
2.5\,d$^{-1}$) with $T_{\rm eff} > 18000$\,K in the {\it TESS} field. 

The median amplitude for all 308 SPB stars is 1.3\,ppt, which is
considerably smaller than for $\beta$~Cep stars.

From ground-based observations, the SPB stars have been found to be slow 
rotators ($\langle v\sin i\rangle = 60 \pm 10$\,km\,s$^{-1}$, \citealt{Balona2009}).  The 
mean projected rotational velocity for 121 {\it TESS} SPB stars with
$10000 < T_{\rm eff} < 18000$\,K is $\langle v\sin i\rangle = 147 \pm 9$\,km\,s$^{-1}$, 
while for all main sequence stars in the same temperature range  
$\langle v\sin i\rangle = 138 \pm 2$\,km\,s$^{-1}$ (1672 stars).  For hotter SPB stars 
$\langle v\sin i\rangle = 148 \pm 12$\,km\,s$^{-1}$ (61 stars), while for main-sequence
stars in the same $T_{\rm eff}$ range $\langle v\sin i\rangle = 140 \pm 2$\,km\,s$^{-1}$ 
(2545 stars).  Clearly, the rotation rate of SPB stars is the same as for 
main sequence stars in the same $T_{\rm eff}$ range as shown in 
Fig.\,\ref{vdist}.

\begin{table}
\caption{List of Maia stars with known projected rotational velocities
(km\,s$^{-1}$).  The effective temperature, $T_{\rm eff}$ (K), and the
priority level (Pr) is given.  The priority codes are as follows:  
1 - spectroscopic modeling; 2 - narrow-band photometry; 3 - UBV photometry; 
4 - BV photometry; 5 - Spectral type.  The stellar luminosity, $\log
L/L_\odot$, is derived from the Gaia DR2 parallax. The projected rotational
velocity, $v\sin i$ (km\,s$^{-1}$), and the  spectral type is also listed.}
\label{maiarot}
\begin{center}
\begin{tabular}{rlrrrl}
\hline
\multicolumn{1}{c}{TIC} & 
\multicolumn{1}{c}{$T_{\rm eff}$} & 
\multicolumn{1}{c}{Pr} & 
\multicolumn{1}{c}{$\log\tfrac{L}{L_\odot}$} & 
\multicolumn{1}{c}{$v\sin i$} & 
\multicolumn{1}{l}{Sp. Type}  \\
\hline
   9887122 &   11324 &  1 &   2.38 & 150.0 &  B6(V)         \\
  12339621 &   10363 &  1 &   2.27 &  52.5 &  B9.5III+      \\
  25425217 &   10404 &  2 &   2.29 & 230.0 &  B9III         \\
  31921921 &   16575 &  2 &   2.84 &  75.0 &  B1I/IIIk      \\
  36557487 &   13922 &  2 &   2.84 & 255.7 &  B8IIIn        \\
  52684359 &   14624 &  2 &   2.90 &  52.8 &  B8IV (shell)  \\
  66594335 &   16008 &  1 &   2.94 & 300.0 &  B2IV-V        \\
  71580820 &   17024 &  1 &   3.35 &  55.0 &  B5III         \\
 116273716 &   12910 &  1 &   2.59 & 156.7 &  B5/7III       \\
 129533458 &   12742 &  1 &   2.84 &  69.5 &  B7III         \\
 144028101 &   13520 &  5 &   2.74 & 266.7 &  B8Ve          \\
 144517863 &   12151 &  1 &   2.27 & 290.0 &  B9V           \\
 160704414 &   13183 &  1 &   2.40 & 367.3 &  B7:V:nn       \\
 169551936 &   12439 &  1 &   2.85 & 123.0 &  B8III         \\
 174662768 &   13918 &  2 &   2.66 &  42.0 &  B5Vn          \\
 191142098 &   10864 &  1 &   1.56 & 212.0 &  B8.5Vn        \\
 191493281 &   15915 &  1 &   2.97 & 167.2 &  B3III/V:      \\
 202431888 &   12024 &  1 &   2.46 &  44.3 &  B9IVSi:       \\
 225263020 &   10028 &  2 &   1.51 & 214.0 &  A0/1V         \\
 234887704 &   13520 &  5 &   2.12 & 286.0 &  B8Ve          \\
 235459271 &   11967 &  1 &   2.57 & 170.8 &  B8Vn          \\
 239219717 &   11776 &  1 &   2.17 & 333.8 &  B6III         \\
 241660076 &   11508 &  1 &   2.11 & 249.5 &  B9.5V         \\
 250137613 &   14670 &  1 &   3.11 &  22.8 &  B5IV          \\
 264540595 &   11583 &  2 &   2.06 & 270.0 &  B9.5V         \\
 270219259 &   13520 &  5 &   2.96 & 301.2 &  B8III shell   \\
 271971626 &   11412 &  1 &   2.29 &  17.0 &  B9IV          \\
 277674241 &   10789 &  1 &   1.85 & 308.0 &  B9Vn:         \\
 301100741 &   12000 &  2 &   2.25 &  84.2 &  B9Si          \\ 
 306548914 &   17466 &  2 &   2.95 & 155.7 &  B5V           \\
 308769611 &   10116 &  1 &   1.80 & 172.0 &  A0V           \\
 323246065 &   12214 &  1 &   2.12 & 343.0 &  B9IV          \\
 331268750 &   13552 &  1 &   2.71 &  88.2 &  B6/7 + B7/8   \\
 341040976 &   12900 &  2 &   2.63 &  25.0 &  B9Si          \\
 354793407 &   14900 &  2 &   2.78 &  51.4 &  B8Hewk.Si     \\
 357310008 &   13100 &  2 &   2.46 &  30.0 &  B9Si          \\
 400445441 &   14562 &  2 &   2.74 &  65.0 &  B5/6IV        \\
 408382023 &   17140 &  5 &   3.08 & 272.6 &  B6V(e)        \\
 450935207 &   15975 &  1 &   2.67 & 320.0 &  B6Vn          \\
 469421586 &   13520 &  5 &   2.85 & 205.0 &  B8IV/V        \\
 469906369 &   10069 &  1 &   2.09 & 220.5 &  B9.5IVn       \\
\hline                        
\end{tabular}
\end{center}
\end{table}

\section{Maia stars}

Perhaps the most interesting result is the presence of a significant number of 
stars with high frequencies cooler than the red edge of the $\beta$~Cep region.
This confirms the detection of such stars by {\it CoRoT}
\citep{Degroote2009b}.  If these are rapidly-rotating $\beta$~Cep or SPB stars,
then all of these Maia variables should have $v\sin i$ considerably larger than
main sequence stars in the same $T_{\rm eff}$ range.  Projected 
rotational velocities are available for 41 of the 131 Maia stars 
(Table\,\ref{maiarot}).  The mean projected rotational velocity is 
$\langle v\sin i\rangle = 173 \pm 17$\,km\,s$^{-1}$, while the for main 
sequence stars in the same temperature range $\langle v\sin i\rangle = 138 \pm 
3$\,km\,s$^{-1}$ from 1672 stars.  There is a difference of two
standard deviations between the two values, which is not considered 
statistically significant.  The frequency distribution is shown in
 Fig.\,\ref{vdist}.

Clearly, Maia stars are not rapidly rotating $\beta$~Cep or SPB stars.  They 
could possibly be composite objects consisting of a non-pulsating B star and 
a $\delta$~Sct star.  Another possible explanation is that 
the effective temperatures are in error and that Maia stars are actually 
$\delta$~Sct variables. The most reliable measure of $T_{\rm eff}$ has been
used.  As can be seen from Table\,\ref{maiarot}, a large proportion of 
stars have $T_{\rm eff}$ measured by modelling the spectrum, so this 
explanation is unlikely.

The median amplitude for all 131 Maia stars is only 0.36\,ppt, which
perhaps explains why these stars were never confirmed to exist from
ground-based observations.

The distribution of high frequencies in Maia variables resembles that in 
$\delta$~Sct stars rather than in $\beta$~Cep variables.  Whereas, dominant 
frequencies as high as 20\,d$^{-1}$ or more are common among the Maia 
variables, the frequency of highest amplitude in BCEP variables rarely exceeds 
10\,d$^{-1}$.  Of the Maia stars with dominant frequencies higher than 
20\,d$^{-1}$, 10\,percent have values of $T_{\rm eff}$ estimated from 
spectroscopy and 20\,percent from spectroscopy and narrow-band photometry.  
All have spectral classifications of A0 and earlier.

\section{Fraction of B stars that pulsate}

\begin{table}
\caption{Number of $\beta$~Cep, SPB and Maia stars, $N_{\rm var}$, and total 
number of main sequence stars, $N_{\rm tot}$,  within a given effective 
temperature range.  The last column gives the percentage of pulsating variables
relative to the total number of main sequence stars.}
\label{stats}
\begin{center}
\begin{tabular}{lrrrr}
\hline
\multicolumn{1}{c}{Type} & 
\multicolumn{1}{c}{$T_{\rm eff}$ range} & 
\multicolumn{1}{c}{$N_{\rm var}$} &
\multicolumn{1}{c}{$N_{\rm tot}$} &
\multicolumn{1}{c}{Percent} \\ 
\hline
BCEP  & 18000--35000 &  284 & 1242 & 22.9 \\
SPB   & 18000--35000 &  162 & 1242 & 13.0 \\
SPB   & 10000--35000 &  289 & 2605 & 11.1 \\
SPB   & 10000--18000 &  127 & 1363 &  9.3 \\
MAIA  & 10000--18000 &   91 & 1363 &  6.7 \\
\hline                        
\end{tabular}
\end{center}
\end{table}

Table\,\ref{stats} shows the number of $\beta$~Cep, SPB and Maia stars in 
different ranges of effective temperature as well as the total number of main 
sequence stars in the same range observed by {\it TESS}.  It is clear that 
pulsation among B stars is not very common.  While it is possible that 
pulsations below the detection level may be present in all B stars, it is likely 
that pulsational instability is confined to only certain stars for reasons 
unknown.  The same result applies for $\delta$~Sct stars \citep{Balona2011g,
Balona2018c, Murphy2019}.  This poses a challenge for current models which do 
not consider possibly small differences in the composition, structure, 
magnetism, rotation etc in the outer layers.  Even if two stars have the same 
effective temperature, luminosity, global abundance and projected rotational 
velocity, these differences may affect pulsational stability and mode 
selection.

\section{Relationship with $\delta$~Scuti and $\gamma$~Doradus stars}

\begin{figure}
\begin{center}
\includegraphics{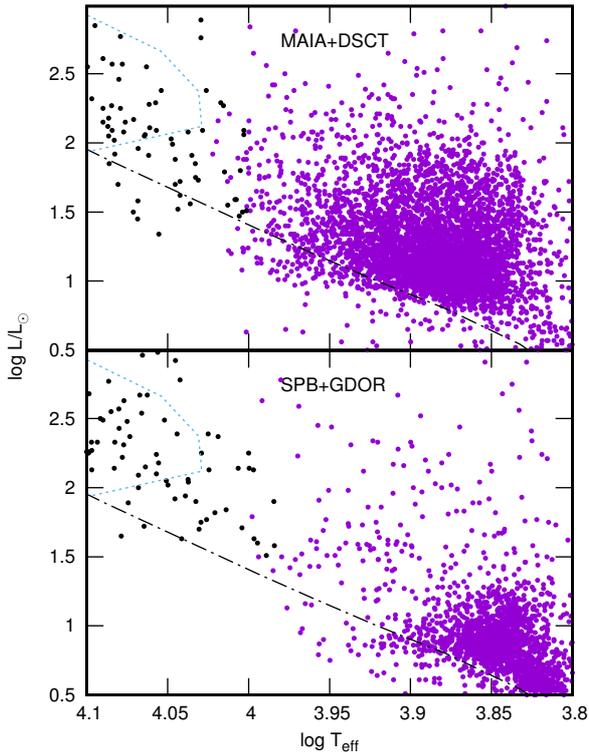}
\caption{The top panel shows the Maia stars (left of $\log T_{\rm eff} =
4.0$) and the DSCT stars (to the right of  $\log T_{\rm eff} = 4$).  Some
DSCT stars hotter than 10000\,K have spectral types indicating cooler
temperatures.  Also shown is the ZAMS (dash-dotted lines) and the theoretical 
cool edge of the SPB stars (dotted line).  The bottom panel shows the same for 
the SPB and GDOR stars observed by {\it TESS}.}  
\label{astars}
\end{center}
\end{figure}

The $\delta$~Sct  and $\gamma$~Dor variables \citep{Antoci2019} seem to be 
driven by the same mechanism and distinguished only by  different mode 
selection \citep{Balona2018c}.  However, they clearly occupy well-defined regions in 
the H--R diagram as shown in Fig.\,\ref{astars}, even though the
classification is based on the presence or absence of peaks with frequencies
in excess of 5\,d$^{-1}$, which is independent of its effective temperature
or luminosity.  From this point of view, it is still useful to distinguish 
between $\gamma$~Dor and $\delta$~Sct stars, even though low frequencies are 
present in both types of variable.

The same is not true for $\beta$~Cep and Maia stars.  Both are characterised by
the presence of high frequencies and, if considered as a single group,
would occupy the whole B star range.  The imposed boundary of 18000\,K between 
the two groups is purely arbitrary, guided by the fact that some correspondence
between the models and observations needs to be made.  In other words, unlike 
the $\delta$~Sct and $\gamma$~Dor stars, there is no information in
the light curve or periodogram which relates to the location of the star in
the H-R diagram.

In Fig.\,\ref{miglio} it can be seen that stars that correspond to models of
$\beta$~Cep variables can be distinguished not only by the presence of high
frequencies, but also by higher pulsation amplitudes.  However,
classification using amplitudes also requires an arbitrary choice of
amplitude.  Moreover, a low amplitude does not distinguish Maia from
$\beta$~Cep because there are many $\beta$~Cep stars with low amplitudes 
similar to Maia variables.  In other words, there is no observational criterion that offers
a clean distinction between $\beta$~Cep and Maia variables and which is 
independent of spectral type.

Likewise, there is no distinct instability strip for SPB stars. Stars with low 
frequencies occur right across the whole B-star main sequence as well as in
blue supergiants \citep{Bowman2019a}.  Clearly, there is a problem with current 
pulsation models which need to be refined.

This study is about pulsation among the B stars and, as a consequence, it
was initially restricted to stars with $T_{\rm eff} > 10000$\,K.  However,
in Fig.\,\ref{hrdiag} the numbers of SPB and Maia variables do not decrease
with decreasing $T_{\rm eff}$ which suggests that both these variables
should be present among the early A stars.  This is indeed the case, as can
be seen in Fig.\,\ref{astars}.  In this figure we show the locations of over
5000 $\delta$~Sct and over 2300 $\gamma$~Dor variables in the H-R diagram from 
{\it TESS} sectors 1--18.  Also included are 113 roAp stars 
\citep{Cunha2019}, of which 68 are new candidates.  These are included for 
completeness.

It should be noted that there are a considerable number of stars classified
as DSCT, even though they are hotter than the observed $\delta$~Sct blue edge 
at about $\log T_{\rm eff} \approx 3.95$.  It is not possible to distinguish
between these hot $\delta$~Sct stars and the Maia variables.  An artificial
boundary of 10000\,K was chosen to separate MAIA and DSCT variables.
 
Similarly, stars classified as GDOR (i.e. frequency mostly less than 
5\,d$^{-1}$) seem to be present which are much hotter than the observed blue 
edge of the $\gamma$~Dor stars at about 7500\,K \citep{Balona2016c}.  These 
hot $\gamma$~Dor stars cannot be distinguished from the SPB stars.  Once
again, a purely arbitrary boundary of 10000\,K was chosen between SPB and
GDOR stars.  The Maia stars seem to be a continuation of the $\delta$~Sct 
variables which then merge into the $\beta$~Cep stars.  The $\gamma$~Dor stars,
which cannot really be considered as separate from the $\delta$~Sct
variables, seem to  merge into the SPB stars.  Of course, this may be 
a result of incorrect effective temperatures, and further research is required.

\begin{figure}
\begin{center}
\includegraphics{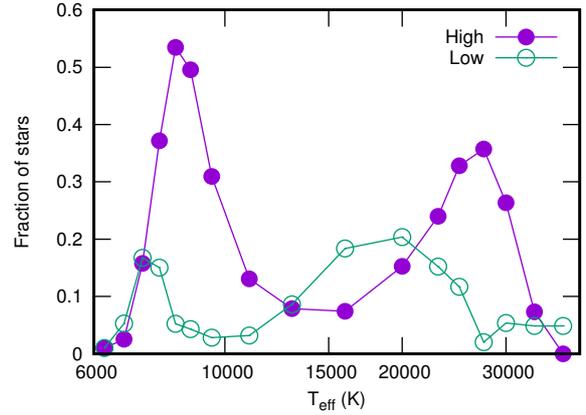}
\caption{The fraction of pulsating stars with high frequencies (i.e.
$\delta$~Sct, Maia and $\beta$~Cep, filled circles) and with low frequencies 
($\gamma$~Dor and SPB, open circles) as a function of effective temperature.}
\label{very}
\end{center}
\end{figure}

In Fig.\,\ref{very} the fraction of high-frequency pulsating stars 
($\delta$~Sct, Maia and $\beta$~Cep variables) and of low-frequency pulsating 
stars ($\gamma$~Dor and SPB variables) is shown as a function of $T_{\rm eff}$.
Even at a minimum of $T_{\rm eff} \approx 15000$\,K, stars pulsating at high
frequencies still comprise about 8\,percent of the total number of main
sequence stars in the same temperature range.  The distinct separation in
effective temperature between DSCT and BCEP stars that non-rotating models 
predict is not seen; instead there are main-sequence stars with similar 
frequencies between the two groups.  The same is true of the GDOR and SPB 
stars, though the situation is less clear in this case.

\section{Conclusions}

Light curves and periodograms of over 50000 stars in {\it TESS} sectors 1--18 
were visually examined.  Where appropriate, each star was assigned a 
variability type.   Using the TIC effective temperatures, many stars 
classified as DSCT ($\delta$~Scuti) or GDOR ($\gamma$~Doradus) were later 
re-classified as BCEP ($\beta$~Cephei), MAIA or SPB based on more reliable 
estimates of $T_{\rm eff}$ and/or their spectral types.  The Maia class is 
necessary because many B stars with high frequencies are too cool to be 
classified as $\beta$~Cep variables.

With the period--effective temperature diagram as a guideline, it became clear 
that a stricter definition of the different types of B-type variables is
required.  For stars hotter than about 18000\,K, a frequency of 2.5\,d$^{-1}$ 
was chosen to distinguish between BCEP and SPB stars. For cooler stars a
frequency of 5\,d$^{-1}$ was chosen as the boundary between Maia and SPB
stars. Using this classification, 327 $\beta$~Cep, 131 Maia and 308 SPB stars 
were detected.  These results confirm the {\it CoRoT} detections of Maia 
variables and SPB stars extending well into the $\beta$~Cep region 
\citep{Degroote2009b}.

In estimating $T_{\rm eff}$ we used a system of priorities where the most
reliable estimates (priority 1) are those where $T_{\rm eff}$ is derived by
modelling the stellar spectrum.  The least precise estimate (priority 5)
uses the spectral type and luminosity class.  The effective temperature for
B stars is still poorly known and badly neglected because modern CCDs are
not very sensitive in the UV range.

It seems that all three groups of pulsating variables have projected
rotational velocities similar to non-pulsating stars in the same temperature
range.  In particular, the Maia stars rotate no faster than other main
sequence stars in the same $T_{\rm eff}$ range.  Therefore this anomalous 
group cannot be explained as rapidly-rotating
$\beta$~Cep or SPB stars.  Because many Maia stars have values of $T_{\rm
eff}$ derived from spectrum modelling, it is also unlikely that they are a
result of erroneous temperature estimates. 

If the pulsating A stars are considered, there appears to be no distinct
grouping of high-frequency pulsators.  The $\delta$~Sct, Maia and
$\beta$~Cep stars seem to merge smoothly with each other.  The $\delta$~Sct 
and $\beta$~Cep groups form two distinct maxima in the relative population and 
also have amplitudes significantly larger than other pulsating stars, but they 
are not isolated groups.  The relatively large number of stars of high 
frequencies outside the traditional instability strips of $\delta$~Sct and 
$\beta$~Cep stars remains unexplained.

The $\beta$~Cep stars observed by {\it TESS} have frequencies which span a
relatively narrow range, mostly between 3--10\,d$^{-1}$.  Their amplitudes
are relatively large compared to the SPB and Maia stars.  A significant 
fraction of the $\beta$~Cep stars pulsate in only a single dominant frequency 
which, by matching the dimensionless frequency with models, may be the second 
radial ($l=0$) or quadrupole ($l=2$) overtone.  These stars have high
amplitudes and low rotation rates.  Why a particular mode with these
properties should be selected is not clear.

Low-frequency pulsations seem to be present across the whole main sequence
from the $\gamma$~Dor stars to the domain of the $\beta$~Cep variables.  In
fact, a substantial number of SPB stars with no high frequencies and $T_{\rm
eff} > 18000$\,K have been found, even though these are not predicted. 
These were first detected by {\it CoRoT} \citep{Degroote2009b}.  Their rotation 
rates are, however, quite normal.  Although there are relatively few of the 
anomalous ``hot $\gamma$~Dor'' stars, they merit further study.

The Be stars are not considered here.  There is a widespread opinion that
the mass-loss mechanism is a result of nonradial pulsation coupled with
near-critical rotation \citep{Rivinius2013}.  Recent results from {\it TESS} 
do not support this conclusion \citep{Balona2019d} since the variations are 
incoherent, though quasi-periodic.  Recent analyses of the projected rotational
velocities also indicate that they do not rotate at near-critical velocity
\citep{Cranmer2005, Zorec2016}. 

These results, though different from what may have been expected based on
current knowledge, are not entirely surprising.  For some time it has been
clear that the simple picture of static radiative envelopes in A and B stars
is in need of revision \citep{Balona2012c, Balona2013c, Balona2016a,
Balona2017a}.  The detection of rotational modulation among a large fraction
of B stars \citep{Balona2019c} shows that current concepts of B star
atmospheres are not compatible with observations.   It seems that the multitude 
of questions relating to pulsation among the A and B stars needs to await a
better understanding of the physics of the upper envelopes of these stars.

\section*{Acknowledgments}

LAB wishes to thank the National Research Foundation of South Africa for 
financial support. Discussions with Dr Peter De Cat, Dr Gerald Handler and Dr 
Keivan Stassun are also gratefully acknowledged.

Funding for the {\it TESS} mission is provided by the NASA 
Explorer Program. Funding for the {\it TESS} Asteroseismic Science Operations 
Centre is provided by the Danish National Research Foundation (Grant agreement 
no.: DNRF106), ESA PRODEX (PEA 4000119301) and Stellar Astrophysics Centre 
(SAC) at Aarhus University. 

This work has made use of data from the European Space Agency (ESA) mission 
Gaia, processed by the Gaia Data Processing and Analysis Consortium (DPAC).
Funding for the DPAC has been provided by national institutions, in particular 
the institutions participating in the Gaia Multilateral Agreement.  

This research has made use of the SIMBAD database, operated at CDS, 
Strasbourg, France.  Data were obtained from the Mikulski Archive for Space 
Telescopes (MAST).  STScI is operated by the Association of Universities for 
Research in Astronomy, Inc., under NASA contract NAS5-2655.

\bibliographystyle{mnras}
\bibliography{maia}

\label{lastpage}

\end{document}